\documentclass[final, 5p,twocolom, times]{elsarticle}

\usepackage{amsmath}
\usepackage{graphicx}
\usepackage{amssymb}
\usepackage{bm}
\biboptions{sort&compress}

\begin{document}

\begin{frontmatter}

\date{Submitted 2 November 2010}

\title{%
Impurity effect in multi-orbital, sign-reversing
s-wave superconductors}
\author[ccse,crest,trip]{Y. Nagai\corref{cor1}}

\author[electro,trip]{K. Kuroki}
\author[ccse,crest,trip]{M. Machida}
\author[tokyo,trip]{H. Aoki}

\address[ccse]{CCSE, Japan Atomic Energy Agency, 6--9--3 Higashi-Ueno,
Taito-ku, Tokyo 110--0015, Japan}
\address[electro]{Department of Applied Physics and Chemistry, The University of Electro -Communications, Chofu, Tokyo 182-8585, Japan}
\address[tokyo]{Department of Physics, University of Tokyo, Hongo, 
Tokyo 113-0033, Japan}
\address[crest]{CREST (JST), 4--1--8 Honcho, Kawaguchi, Saitama 332--0012, Japan}
\address[trip]{TRIP (JST), 5 Sanbancho Chiyoda-ku, Tokyo 102-0075, Japan}

\begin{abstract}
We study the impurity effects on the transition temperature $T_{c}$ with use of the $T$-matrix approximation. 
We propose a way to visualize the multi-orbital effect by introducing the hybridization function characterizing the multi-orbital effect for the impurity scattering.  
Characterizing function does not depend on the superconducting pairing symmetry, since this function is defined by the eigenvectors in normal states. 
The result indicates that an impurity-robust superconductivity does not necessarily imply a sign-preserving pairing. 
Visualizing the hybridization effect in the effective five-band model for LaFeAsO, 
we show that the impurity effect on $T_{c}$ is relatively weaker than that in single-band models.

\end{abstract}

\begin{keyword}
Impurity effects
\sep 
Iron-based superconductors
\sep 
multi-orbital systems
74.20.Rp \sep 74.70.Xa
\end{keyword}

\end{frontmatter}

\section{Introduction}

The discovery of the iron-based superconductor\cite{Kamihara}, 
which is a five-orbital system with three bands involved in the gap function\cite{KurokiPRL}., is stimulating renewed interests in multi-orbital superconductivity.  
Specifically, a sign-reversing s-wave pairing (s$\pm$), which exploits the multi-orbital nature of the system, has been theoretically proposed as a candidate\cite{Mazin,KurokiPRL}.   
While various experimental measurements for 
determining the pairing symmetry in the iron-based superconductors 
are accumulating, which include 
the penetration depth, thermal conductivity, 
ARPES, STM, and NMR, 
one important test is the impurity effect on superconducting transition temperature $T_{c}$.   This has in fact been studied intensively both experimentally and theoretically \cite{Li,Guo,Sato,Onari}. 
There are two trends found in the experiments on the non-magnetic impurity effect on $T_{c}$ for LaFe$_{1-x}$Zn$_{x}$AsO. 
The experiment by Li {\it et al.} suggests that 
the doping of Zn in LaFeAsO$_{1-y}$F$_{y}$ does not reduce $T_{c}$\cite{Li}. 
The result by Guo {\it et al.} suggests that $T_{c}$ decreases significantly 
for a minimal level of Zn doping in LaFeAsO$_{0.85}$\cite{Guo}. 
A theoretical work by Onari and Kontani suggests that a fully gapped sign-reversing $s$-wave state is 
very fragile against non-magnetic impurities\cite{Onari}. 
On the other hand, the compound containing phosphorus, LaFePO, whose gap function has line-nodes, is robust against non-magnetic \\impurities\cite{Tajima}.  
With this background, here 
we theoretically study the problem from a broader context, i.e., we want to identify which factors are crucial in determining $T_{c}$ in dirty, multi-orbital superconductors, especially for the sign-reversing s-wave.

\section{Formulation}

The self-energy in the $T$-matrix approximation in the orbital-representation is expressed as 
$\check{\Sigma}^{\rm orbital} = n_{\rm imp} \check{T}$, where 
$n_{\rm imp}$ is the density of impurities, and 
$\check{T}  = (\check{1} - \check{V} \check{G}_{\rm loc} )^{-1} \check{V}$.  
Here $\check{V} = \hat{V} \check{\sigma}_{z}$, and 
the local Green's function $\check{G}_{\rm loc}$ is defined as $\check{G}_{\rm loc} =\frac{1}{N} \sum_{\bm q} \check{G}^{\rm orbital}({\bm q})$ with 
\begin{align}
\check{G}^{\rm orbital}({\bm q},i \omega_{n})  &\equiv 
\left(\begin{array}{cc}
\hat{G}^{\rm orbital}({\bm q}, i \omega_{n}) & -\hat{F}^{\rm orbital}({\bm q}, i \omega_{n}) \\
-\hat{F}^{\rm orbital\dagger}({\bm q}, i \omega_{n}) & - \hat{G}^{\rm orbital}({\bm q},- i \omega_{n})
\end{array}\right).
\end{align}
Throughout the paper, $\hat{a}$ denotes an $n \times n$ matrix in the orbital space while $\check{a}$ a $2n \times 2n$ matrix composed of the $2 \times 2$ Nambu space and the $n \times  n$ orbital space. 
With use of the unitary matrix $\check{P}_{{\bm k}}$ that diagonalizes the Hamiltonian in the orbital basis, the self-energy in the band representation is 
expressed as 
$\check{\Sigma}_{\bm k}^{\rm band}  = n_{\rm imp} [\check{1} - \check{V}^{\rm band}({\bm k}) \check{G}_{\rm loc}^{\rm band}({\bm k})]^{-1} \check{V}^{\rm band}({\bm k})$ with $\check{V}^{\rm band}({\bm k}) \equiv \check{P}_{\bm k}^{\dagger} \check{V} \check{P}_{\bm k}$. 
Considering the orbital-indepent impurity potential $\check{V} = V_{0} \check{\sigma}_{z}$ introduced in Ref. \cite{Onari},  the self-energy is expressed as 
$\check{\Sigma}_{\bm k}^{\rm band} = n_{\rm imp} [\check{1} - V_{0} \check{\sigma}_{z}  \check{G}_{\rm loc}^{\rm band}({\bm k})]^{-1} V_{0} \check{\sigma}_{z}$.
We then obtain the diagonal elements of the normal part of the self-energy 
as 
\begin{align}
(\hat{\Sigma}_{\bm k}^{ { \rm band, N}})_{ii} = V_{0} + V_{0}^{2} (\hat{G}_{\rm loc}^{{\rm band}})_{ii}({\bm k}) + \cdots, \label{eq:sigma}
\end{align}
with 
\begin{align}
(\hat{G}_{\rm loc}^{\rm band})_{ii}({\bm k}) &= \frac{1}{N} \sum_{{\bm q}} \sum_{l} |C_{li}({\bm q},{\bm k})|^{2} G_{ll}^{\rm band}({\bm q}). \label{eq:gloc}
\end{align}
Here we have introduced a unitary matrix $\hat{C}({\bm q},{\bm k})$ whose element $C_{ij}({\bm q},{\bm k})$ can be written as 
\begin{align}
C_{ij}({\bm q},{\bm k}) = \vec{p}_{{\bm q} i}^{\dagger} \vec{p}_{{\bm k} j}, \label{eq:c}
\end{align}
where $\vec{p}_{{\bm k} j}$ denotes the $j$-th eigenvector of the Hamiltonian with momentum ${\bm k}$ in the normal state, while ${\bm k}$ and ${\bm q}$ denote, 
respectively, the initial- and final-state quasiparticle momenta in impurity scatterings. 
Equations ~(\ref{eq:sigma}) and (\ref{eq:gloc}) imply that 
the $l$-th band hybridization affects the $i$-th band self-energy $(\hat{\Sigma}^{\rm band, N}_{\bm k})_{ii}$ through 
$|C_{li}({\bm q},{\bm k})|^{2} G_{ll}^{\rm band}({\bm q})$. 
Hence the factor $|C_{li}({\bm q},{\bm k})|^{2}$ characterizes the multi-orbital effect for the impurity scattering. 
From Eq.~(\ref{eq:c}) we find that the more the eigenvectors with initial-state momentum ${\bm k}$ and final-state momentum ${\bm q}$ resemble 
with each other, the stronger the inter-band impurity effect on $T_{c}$ becomes.  

Thus a speciality of a multi-band system 
is that, on top of the band dispersion and Fermi surface, the 
character of eigenvectors acts as an important factor determining the 
way in which the impurity effect appears. 
We can thus call the factor $|C_{li}({\bm q},{\bm k})|^{2}$ the {\it impurity scattering intensity}. 
We can use this property to construct a model that is robust 
against the impurity effect. 
For instance, the impurity effect does not appear on $T_{c}$ in a two-orbital model described by $H^{ii}_{\bm k} = - [\cos(k_{x}) + \cos(k_{y})], 
H^{ij}_{\bm k} = t'$, since 
the impurity scattering intensity in this case reduces to 
$|C_{li}({\bm q},{\bm k})|^{2} = \delta_{li}$ everywhere in momentum space. 

\section{Visualized hybridization effect on the impurity scattering}
The general scheme above enables us to 
visualize the hybridization effect on the impurity scattering. 
In doing so, we can concentrate on 
the Fermi momentum, since the Green function has amplitudes 
localized around the Fermi energy in Eq.~(\ref{eq:gloc}). 
Then $|C_{li}({\bm q},{\bm k})|^{2}$ can be parameterized 
as $|C_{li}(\theta_{l}, \theta_{i})|^{2}$, where 
$\theta_{l(i)}$ describes the position of the final- (initial-) state on the $l$-th ($i$-th) Fermi surface. 
Let us visualize the hybridization effect in the effective five-band model proposed by  Kuroki {\it et al.} for LaFeAsO \cite{KurokiPRL}. 
In this model with the Fermi energy $E_{\rm F} = 10.94$eV, there are two hole Fermi pockets around $(k_{x}, k_{y}) = (0,0)$, and 
two electron pockets around $(0, \pi)$ and $(\pi,0)$ as displayed in Fig.~1.

First, we visualize the {\it inter-band} impurity scattering intensity of the quasiparticles between the hole and electron Fermi surfaces.  
Here, we introduce the band-index $i$ whose energy $\epsilon_{i}$ satisfies the relation $\epsilon_{i} > \epsilon_{j}$ $(i > j)$.  
The hole Fermi surfaces on the $2$nd and $3$rd bands are mainly constructed from $d_{xz}$ and $d_{yz}$ orbitals. 
The dominant component of the eigenvectors at the electron Fermi surface on the $4$th band depends on the Fermi wave-number. 
In Fig.~3
, we show the dominant components in the orbital basis on each Fermi surface. 
In the case of the impurity scattering between $2$nd and $4$th bands, 
the intensity strongly depends on both of the initial-state $\theta_{2}$ and the final-state $\theta_{4}$ as shown in 
Fig.~2(a). 
For example, the intensity at $(\theta_{2},\theta_{4}) = (3 \pi/2,\pi/2)$ is almost zero, so that the inter-band impurity scatterings 
between $\theta_{2} = 3 \pi/2$ and $\theta_{4} = \pi/2$ do not affect the superconducting transition temperature $T_{c}$ even if 
these are sing-reversing scatterings. 
The intensity at  $(\theta_{2},\theta_{4}) = (\pi/2,\pi)$ as shown in Figs.~2
(a) and  
3
 has a maximum value 
$|C_{24}(\theta_{2}, \theta_{4})|^{2} \simeq 0.6$.
By contrast, the impurity scattering between $3$rd and $4$th bands does not affect $T_{c}$ regardless of the initial and final momenta, 
since the intensity is small everywhere as shown in Fig.~2(b).

Second, we turn to the {\it intra-band} impurity scattering intensity $|C_{22}(\theta_{2},\theta_{2}')|^{2}$ and $|C_{33}(\theta_{3},\theta_{3}')|^{2}$. 
As shown in Fig.
~4
, the intensities on a same band are bigger than that between different bands on the whole.
On the line satisfying the relation $\theta_{i} = \theta_{i}'$, the intensity $|C_{ii}(\theta_{i},\theta_{i} )|^{2}$ become 
$|C_{ii}(\theta_{i},\theta_{i} )|^{2} = 1$, since the eigenvectors are same. 
The intensities $|C_{ii}(\theta_{i},\theta_{i} + \pi)|^{2}$ differ from $|C_{ii}(\theta_{i},\theta_{i})|^{2} = 1$. 
This originates from an absence of the inversion symmetry in the eigenvector $\vec{p}_{{\bm k} j}$.   Although this may first seem strange, 
the Hamiltonian for momentum ${\bm k}$ in the orbital-basis is not equal to that 
for momentum $-{\bm k}$, since 
the positional relation between the iron atoms and the arsenic atoms 
differs between the directions for ${\bm k}$ and $-{\bm k}$. 

From these figures~2
 and 4
we find that the impurity scattering intensities $|C_{li}(\theta_{l},\theta_{i})|^{2}$ in this effective five-band model 
are small on the whole. 
Focusing  the fact that the intensities in single band models are always $|C_{11}(\theta_{1},\theta_{1})|^{2} = 1$, 
these figures suggest that the impurity effect on $T_{c}$ in the effective five-band model is relatively weaker than that in single-band models.

Finally, we show the impurity intensity dependence of $T_{c}$ in Fig.~5
 with use of the self-consistent $T$-matrix approximation in the sign-reversing $s$-wave superconductor 
for an impurity density $n_{\rm imp} = 0.01$.  
The result, which is essentially the same as the result in Ref.~\cite{Onari}, 
indicates that an attractive impurity potential does not significantly 
reduce $T_{c}$. 
This behavior originates, in the present view, from the multi-orbital hybridization effect characterized by $|C_{li}(\theta_{l},\theta_{i})|^{2}$.

\section{Conclusion}

We have studied the impurity effects on $T_{c}$ with use of the $T$-matrix approximation. 
We found that the hybridization function $|C_{li}(q,k)|^{2}$ (Eq.~(\ref{eq:c})) characterizes the multi-orbital effect for the impurity scatterings. 
The more the eigenvectors at initial-state and final-state momenta are similar to each other, 
the stronger the inter-band impurity effect on $T_{c}$ becomes. 
We thus proposed a way to visualize the multi-orbital effect. 
The figures visualizing the multi-orbital effect suggest that the impurity effect on $T_{c}$ in the effective five-band model is relatively weaker than that in single-band models. 
The above results do not depend on the superconducting pairing symmetry, since 
$|C_{li}({\bm q},{\bm k})|^{2}$ is defined only by the eigenvectors in normal states. 
The message of this result is that there can be impurity-robust sign-reversing $s$-wave pairing symmetry in the iron-based superconductors.  
Conversely, an impurity-robust superconductivity does not necessarily 
imply sign-preserving pairing.  
In addition, with use of the present visualization, one might find the paring symmetry with line-nodes where the impurity effects do not appear on $T_{c}$ in such materials as LaFePO.


\newpage

\section*{Figure captions}

\noindent
Figure 1: Fermi surfaces in the effective five-band model at the Fermi energy $E_{\rm F} = 10.94$ eV. \\

\noindent
Figure 2: Inter-band impurity scattering intensity of the quasiparticles between the $2$nd band and the $3$rd band (a) , or 
between $2$nd band and $4$th band (b).\\

\noindent
Figure 3: Dominant character of the eigenvectors is 
schematically shown on the Fermi surfaces. 
The different colors denote different hybridization in the orbital basis. 
Blue dots represent an example of the initial and final momenta. \\

\noindent
Figure 4: Intra-band impurity scattering intensity on the $2$nd band (a), 
or on the $3$rd band (b). \\

\noindent
Figure 5: Impurity-intensity ($I$) dependence of $T_{c}$ in the sign-reversing $s$-wave superconductor.\\

\newpage
\section*{Figures}
\begin{figure}[htbp]
\centering
\scalebox{0.7}[0.7]{\includegraphics{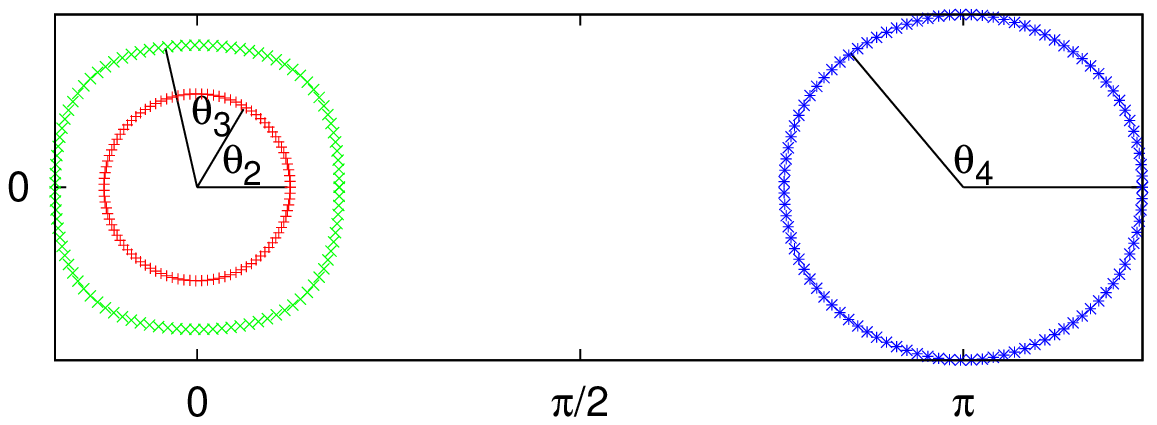}}
\label{fig:fermi} 
\caption{}
\end{figure}

\begin{figure}[htbp]
  \begin{center}
    \begin{tabular}{p{49mm}p{40mm}}
      \resizebox{54mm}{!}{\includegraphics{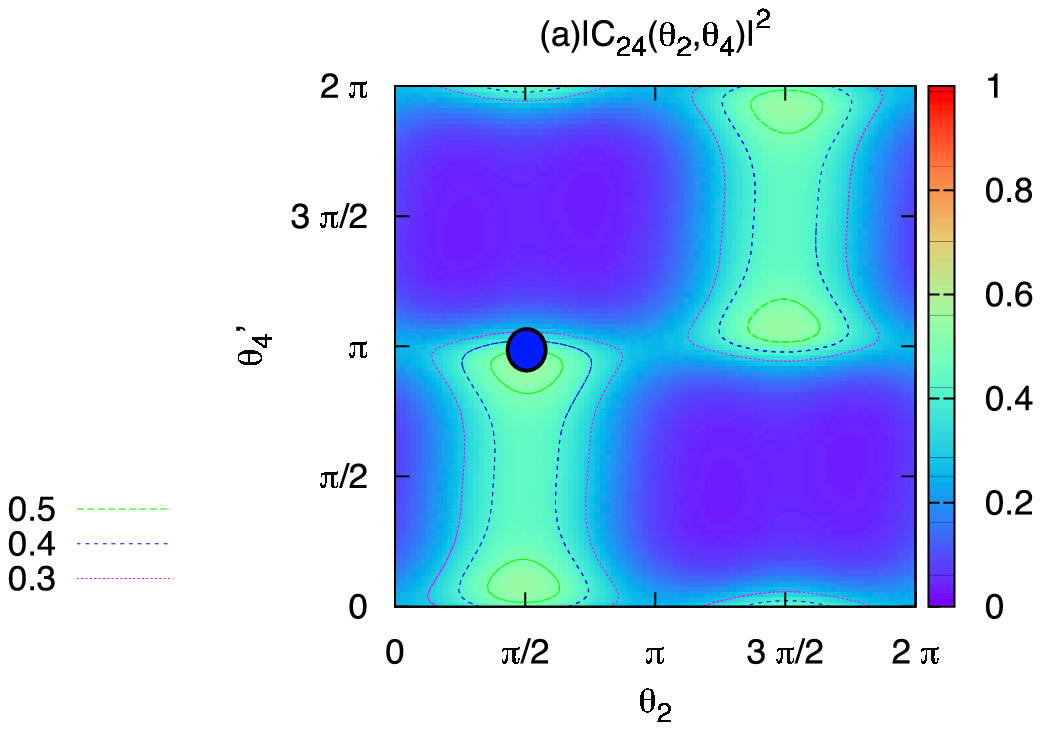}} &
      \resizebox{45mm}{!}{\includegraphics{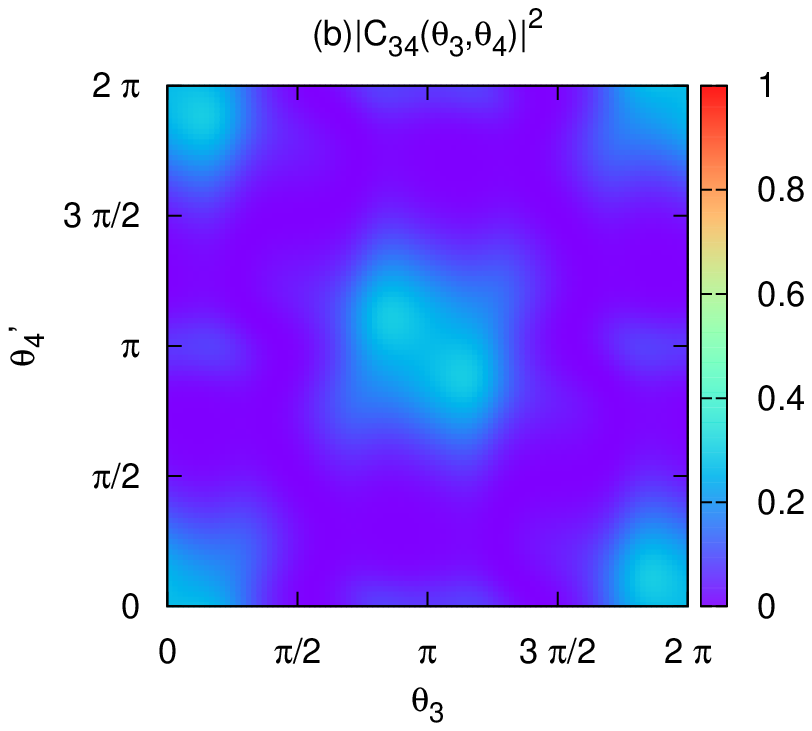}} 
    \end{tabular}
    \label{fig:c24}
    \caption{
}
  \end{center}
\end{figure}

\begin{figure}[htbp]
\centering
\scalebox{0.5}[0.5]{\includegraphics{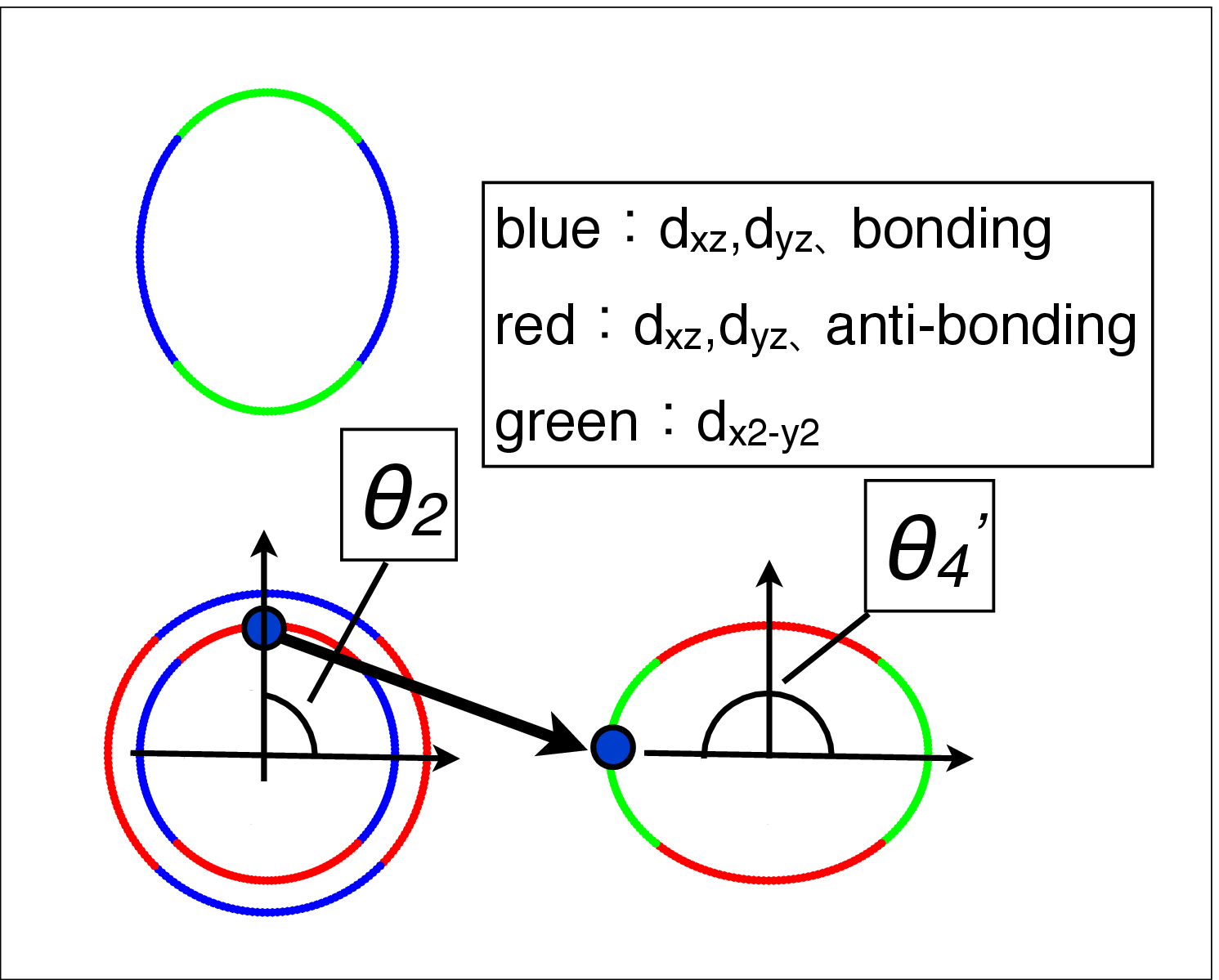}}
\label{fig:bonding} 
\caption{}
\end{figure}

\begin{figure}[htbp]
  \begin{center}
    \begin{tabular}{p{49mm}p{40mm}}
      \resizebox{54mm}{!}{\includegraphics{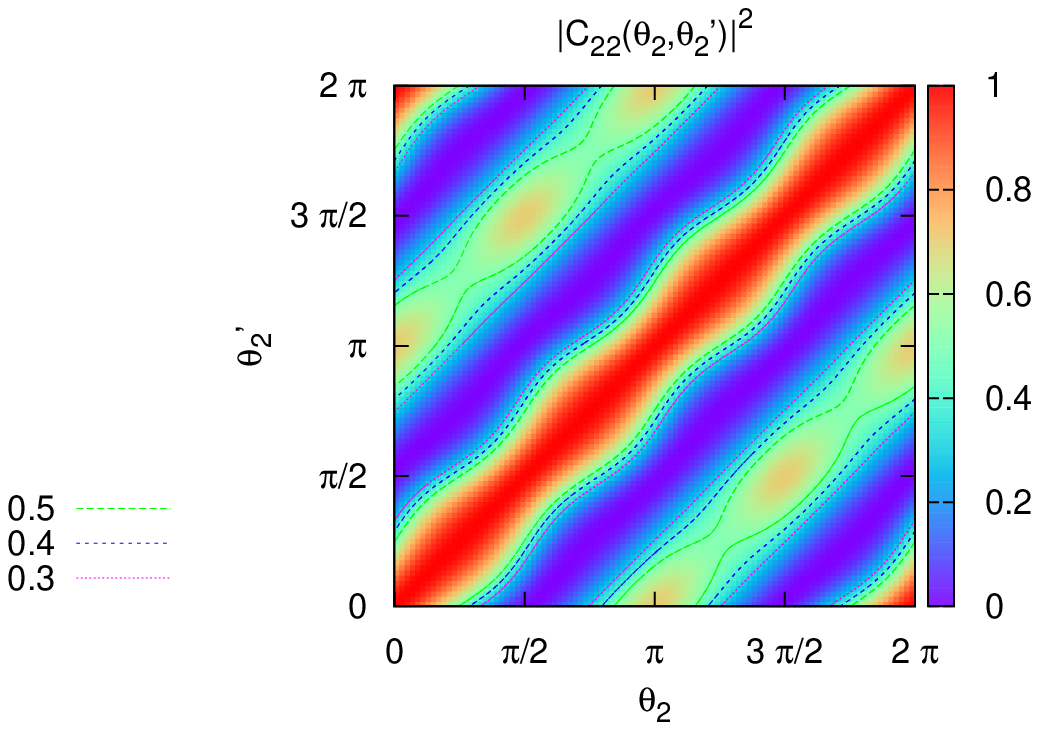}} &
      \resizebox{45mm}{!}{\includegraphics{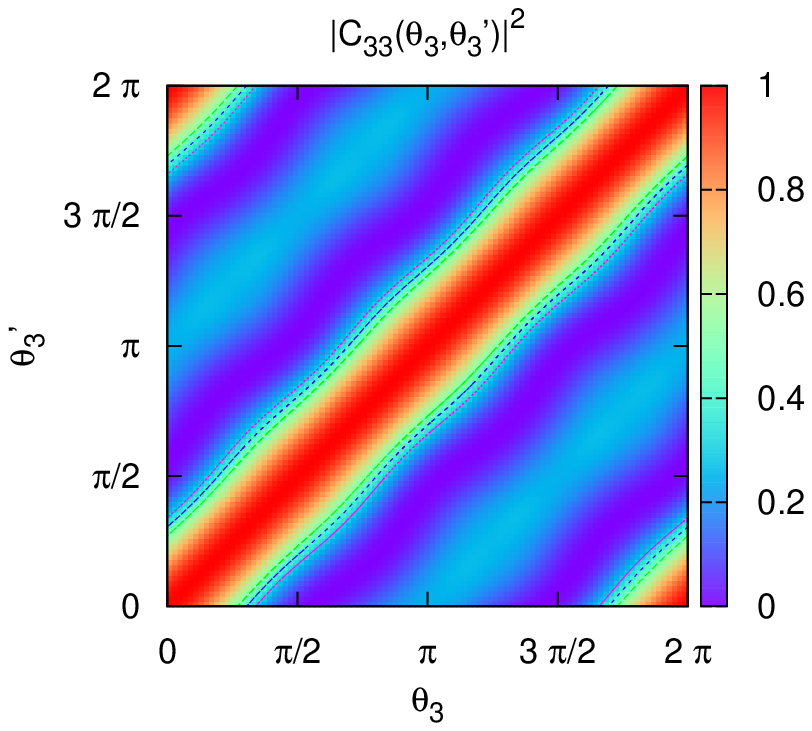}} 
    \end{tabular}
    \label{fig:c22}
\caption{
}
  \end{center}
\end{figure}

\begin{figure}[htbp]
\centering
\scalebox{0.7}[0.7]{\includegraphics{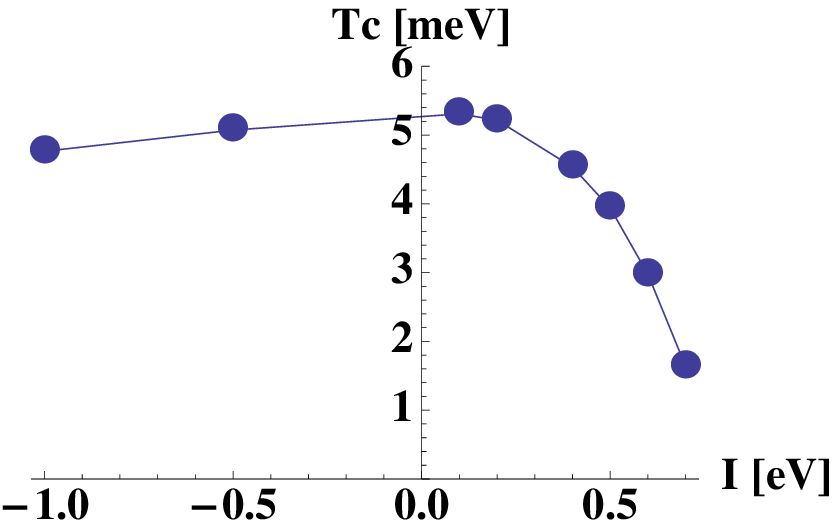}}
\label{fig:impurity} 
\caption{
}
\end{figure}

\end{document}